# LLM-Assisted Abstract Screening with OLIVER: Evaluating Calibration and Single-Model vs Actor–Critic Configurations in Literature Reviews

Authors: Kian Godhwani[1], David Benrimoh[1]

1. McGill University, Department of Psychiatry

Corresponding Author Email: <u>kian.godhwani@mail.mcgill.ca</u>

## Abstract

**Introduction:** Abstract screening is a major bottleneck in systematic and scoping reviews, requiring substantial time, labor, and cost. Recent work suggests large language models (LLMs) can accelerate screening, but prior evaluations focus on earlier LLMs, standardized Cochrane reviews, single-model setups, and accuracy as the primary metric, leaving generalizability, configuration effects, and calibration largely unexamined.

**Methods:** We developed **OLIVER** (**O**ptimized **L**LM-based **I**nclusion and **V**etting **E**ngine fo**r R**eviews)[1], an open-source pipeline for LLM-assisted abstract screening that enables structured criterion evaluation and requires models to output both binary decisions and confidence scores. We evaluated multiple contemporary LLMs across two non-Cochrane systematic reviews differing in size and criterion complexity. Performance was assessed at both the full-text screening and final inclusion stages using sensitivity, specificity, accuracy, AUC, and calibration metrics. We further tested an actor–critic screening framework combining two lightweight models under three aggregation rules: a) any-include, b) require agreement, and c) critic veto configuration.

**Results:** Across individual models, performance varied widely. In the smaller Review 1 (821 abstracts, 63 final includes), several models achieved high sensitivity for final includes but at the cost of substantial false positives and poor calibration. In the larger Review 2 (7741 abstracts, 71 final includes), most models were highly specific but struggled to recover true includes, with prompt design influencing recall. Calibration was consistently weak across single-model configurations despite high overall accuracy. Actor–critic screening improved discrimination and markedly reduced calibration error in both reviews, yielding higher AUCs and substantially lower Brier scores, though gains in sensitivity were modest.

**Discussion:** LLMs may eventually accelerate abstract screening, but single-model performance is highly sensitive to review characteristics, prompting, and calibration is limited. An actor–critic framework improves classification quality and confidence reliability while remaining computationally efficient, enabling large-scale screening at low cost. By jointly evaluating accuracy, configuration effects, and calibration across heterogeneous reviews, this study provides a more rigorous assessment of LLM-assisted screening and establishes OLIVER as a practical, generalizable foundation for exploring integration of LLMs into evidence synthesis workflows. However, poor sensitivity in one of the two reviews studied demonstrates that while LLMs may be a useful tool for assisting with reviews, they are not yet capable of reliably replacing human screeners for complex research questions. Further research on prompting, as well as work to train LLMs on larger swaths of the scientific literature, are needed for further advances.

---

[1] The pipeline can be accessed for free via: <u>https://github.com/kiangodhwani/oliver-llm-abstract-screening</u>

# Introduction

**LLMs for Abstract Screening: Progress and Limitations**

Literature reviews such as systematic or scoping reviews are essential for synthesizing and analyzing evidence within almost every field of research [1]. Reviews allow researchers to understand what has been explored, identify similarities or conflicts in findings across studies, interpret aggregations of results, and identify gaps that need further exploration [1]. A critical part of the review process undertaken after the research question has been formulated and search strategy implemented is the abstract screening process, where abstracts are evaluated against a set of inclusion and exclusion criteria to determine how well they fit within the scope of the research question [1].

The abstract screening process is tedious and labour intensive [1], [2], [3], [4], [5], [6]. In addition, often only 1-10% of abstracts screened are included in the final review [1], [7]. Search strategies often yield several thousand abstracts, each of which take on average 30 seconds to screen for an advanced reviewer [3], [8]. This can cause the abstract screening process to take months with one review reporting that it took 189 days to double-screen 14,923 abstracts [1]. Another study found that the cost of a systematic literature review, driven almost entirely by human labor in processes such as abstract screening and data extraction, could be as high as $141,194 [2].

Given these drawbacks, a variety of attempts have been made to streamline the abstract screening process using machine learning techniques, primarily in the form of text mining [8], [9]. Early tools such as Abstrackr used supervised learning with topic modeling and active learning to predict relevance, sometimes suggesting inclusion or exclusion [8]. Although this reduced workload in some cases, performance was variable and sensitivity could drop when training data were limited [8]. Later tools like Rayyan adopted a more conservative strategy, using Term Frequency-Inverse Document Frequency and continuous active learning to prioritize abstracts rather than classify them [9]. This improved consistency and usability but still required humans to screen most records, highlighting the constraints of traditional text-mining methods [8], [9].

Recent efforts have attempted to use large-language models (LLMs) to speed up the abstract screening process, with promising results [10], [11], [12], [13], [14], [15]. These studies used varying methods, often incorporating different prompt templates and training phases to optimize performance. One study [16] found that optimized prompts that forced step-by-step reasoning against screening criteria achieved better accuracy, sensitivity and specificity compared to zero shot prompts. Another [12] found that prompts biased towards inclusion were, in some cases, able to achieve 100% sensitivity in screening results. Dennstädt et al. (2024) used a simpler approach, inputting structured prompts that contained the title, abstract, and relevant criteria, evaluating the efficacy of 4 open source LLMs in performing abstract screening on 10 published systematic reviews, with models able to achieve high sensitivity, but low specificity.

However, there are some limitations in these findings. For example, Cao et al. (2025) and Sanghera et al. (2025) evaluated LLM's on high-quality Cochrane reviews only; these constituted 3.5% of published reviews between 2013-2022 [17]. This limits the generalizability of their findings to broader or less standardized examples of evidence synthesis [17]. Additionally, earlier work [10], [11], [13] only tested the accuracy of early LLM's like gpt-3.5, gpt-4o, FlanT5 and Mixtral which have been outperformed by newer models on multiple benchmarks [18]. Varying approaches across studies have prevented a clear comparison of performance across models and review types, making it difficult to understand how well current methods generalize beyond varying experimental setups.

**Beyond Single-Model Screening: Ensemble Configurations and Model Calibration**

It is also unclear how effective other configurations of abstract screening, outside of single-model-screening, may be in improving overall accuracy. Sanghera et al. (2025) explored ensemble configurations of abstract screening, including one where two models were made to screen in parallel, with final inclusion requiring agreement between both models, improving recall and precision. These findings suggest value in testing other unexplored configurations that more closely mirror human screening techniques. One such approach worth exploring, which mirrors the conflict resolution process in abstract screening, would be an actor-critic configuration. This would involve one model, the actor, screening the dataset, while the critic goes through the actor's decisions, and provides a decision of their own. Such a configuration also comes with the added benefit of double-screening abstracts, similar to how they are screened by humans.

Another largely unexplored area concerns the calibration of models during abstract screening. Calibration reflects whether a model's stated confidence meaningfully corresponds to the correctness of its decisions, and whether confidence aligns with overall screening performance [19], [20], [21]. This is especially important as recent work suggests well-calibrated recommendations are central to building trust in human–AI decision making workflows [19], [20], [21]. However, assessing calibration is challenging as even the most advanced models exhibit poor out-of-the-box calibration [22], [23]. Furthermore, within the task of abstract screening, models may not apply all inclusion and exclusion criteria consistently across abstracts, and may struggle to produce confidence estimates that accurately reflect the process which actually yielded their decision [13], [22], [23].

**Study Objectives**

To address these limitations, we developed OLIVER (**O**ptimized **L**LM-based **I**nclusion and **V**etting **E**ngine fo**r R**eviews), an open-source pipeline that enables the systematic use of contemporary large language models for abstract screening. Within OLIVER, users can upload a CSV file containing article titles and abstracts, select a specific language model, and input inclusion and exclusion criteria used in the review. From this information, and similar to the approach described by Dennstädt et al. (2024), OLIVER programmatically constructs a structured screening prompt (Appendix) automatically on the backend composed of explicit instructions, user-defined criteria, and a checklist designed to ensure that the model evaluates each abstract against all inclusion and exclusion requirements in a consistent order. The prompt further requires the model to output a binary screening decision (include or exclude), and a confidence score between 0 and 1, enabling downstream analysis of confidence estimates. While our requirement of having the model output a confidence score does not address the implicit challenge LLM models have with out-of-the-box calibration, it presents an opportunity to explore how closely confidence outputs align with model accuracy.

In addition to single-model screening, OLIVER implements an actor–critic pipeline. In this framework, an initial "actor" model independently screens each abstract and outputs a decision, rationale, and confidence score. A separate "critic" model then re-evaluates the same abstract with full access to the original inclusion and exclusion criteria, the actor's decision and confidence, and either endorses or corrects the initial judgment. Final screening decisions are determined using configurable aggregation rules, such as requiring agreement to include or granting the critic veto power. We hypothesized that this actor–critic approach could help potentially identify and correct inconsistent criterion application, mitigate overconfident errors by individual models, and support more stable screening behavior across repeated runs. OLIVER supports multiple state-of-the-art language models, including recent versions of GPT (OpenAI), Claude (Anthropic), Gemini (Google), and Grok (xAI), enabling direct comparison across providers within a consistent screening and evaluation framework. In addition, custom parallel processing scripts were developed for each model to support large-scale screening workloads that exceed standard API rate limits.

In this study, we evaluated OLIVER on two non-Cochrane systematic reviews. One review was already published [24] and one was recently completed but not yet published [25]. Consistent with prior work [12], [13], each model was assessed against the final set of included studies, and model-specific behaviours were compared to identify strengths and weaknesses. Model performance was also assessed against studies that passed the abstract screening phase only, i.e., studies that made it to full text screening, but may or may not have been included by the end of the data extraction process. This was done to better mimic the review workflow.

To our knowledge, this study is the first to evaluate LLM's for abstract screening using a structured actor–critic adjudication framework alongside formal calibration analysis based on model-reported confidence. Together, these contributions provide a more rigorous and generalizable framework for assessing the reliability of LLM-assisted abstract screening in literature reviews.

## Methods

### Reviews

Two reviews were analyzed using OLIVER. Review 1 (821 abstracts) has recently been submitted as a pre-print and explores the integration of social determinants of health in digital phenotyping studies in mental health [25]. Review 2 (7741 abstracts) was a systematic review and meta analysis exploring the proportion of patients who experience a prodromal episode before psychosis onset [24]. Both reviews required double-screening of abstracts with conflicts mediated by the principal investigator [24], [25].

The screening criteria for Review 1 & 2 were complex in differing ways (Appendix). Specifically, Review 1 involved broad and heterogeneous inclusion criteria that used the population-intervention-comparator-outcome (PICO) structure [26] spanning multiple mental health diagnoses, at-risk populations, digital phenotyping modalities, and study designs [25]. In contrast, Review 2 applied narrower inclusion and exclusion criteria, focusing instead on consistent definitions of prodromal symptoms, explicit reporting of prevalence estimates, and exclusion of studies where prodrome was an inclusion criterion [24]. Together, these reviews provided complementary test cases for evaluating LLM-assisted abstract screening under PICO-structured versus non-PICO eligibility criteria.

**Models**

First, we evaluated the performance of individual large language models across both reviews. Models assessed included gpt-4o-mini, gpt-5, gpt-5-mini, gpt-5-nano [27], Claude Haiku, Claude Sonnet [28], grok-4-fast [29], and Gemini Flash [30]. This selection encompassed models with advanced "reasoning" capabilities as well as those optimized for speed and efficiency, which allowed us to capture a broad spectrum of LLM behaviour. The models also varied substantially in computational cost. For example, gpt-5 and Claude Sonnet are expensive and designed for high-accuracy "reasoning" tasks, whereas gpt-5-nano and Claude Haiku are lightweight models that cost only a fraction of the price per million tokens [27], [28]. Including models with distinct cost–performance profiles made it possible to examine how trade-offs between "reasoning" ability, speed, and cost influence screening accuracy. Broad model selection also facilitated characterizing model-specific strengths and limitations in the context of systematic review screening.

Once the accuracy of individual models was established, and building on the work of Sanghera et al. (2025) we sought to explore other screening configurations and applied an actor-critic framework to assess whether combining models improved classification accuracy. We evaluated three configurations: 1) an any-include rule where a record was included if either model proposed inclusion, 2) a critic-veto rule where the actor proposed inclusion but the critic could override it, and 3) a consensus rule similar to the configuration used by Sanghera et al. (2025) requiring both models to independently support inclusion. We evaluated two actor–critic pairings: grok-4-fast as the actor with gpt-5-nano as the critic, and gpt-5-mini as the actor with grok-4-fast as the critic.

**Prompts**

Two prompting strategies were tested: a prompt with unmodified inclusion and exclusion criteria, which was applied to both reviews, and a stratified prompt that organized criteria by population, comparator, outcome, and study characteristics, which was applied only to the second review in an effort to improve performance. This stratification was introduced for the second review because its inclusion criteria were more heterogeneous and, we hypothesized, required clearer distinctions to support consistent model "reasoning".

We began by applying each review's original inclusion and exclusion criteria within a single unified prompt. Review 1 already used stratified criteria organised by the PICO structure (Appendix), and models performed relatively well on this review. In contrast, models performed poorly on Review 2 under its original non-stratified criteria. Based on these results, we restructured Review 2's criteria using the same stratified PICO format. Furthermore, to try and improve performance, the criteria were also tested in a modified format to bias the model towards inclusion, which previous work has shown can improve sensitivity in screening [12]. See the appendix section for further details. Importantly, this modified prompt was tested only on lightweight models (Tables 2 & 4) as they demonstrated strong baseline performance on both reviews, and are the most likely models to be adopted due to their lower computational costs and memory requirements [31].

Once the inclusion and exclusion criteria are inputted on the front-end, a unified screening prompt that presents the criteria alongside the article title and abstract was implemented (see appendix). Within this prompt, the model is instructed to evaluate each abstract against the provided criteria and make an inclusion decision. To reduce variability in criterion application, the prompt appends a structured checklist that requires the model to explicitly assess each inclusion and exclusion criterion in a fixed order. The model is instructed to include an abstract only if all inclusion criteria are met and no exclusion criteria apply, and to exclude the abstract otherwise. The model is required to output a binary decision (include or exclude), and a confidence score between 0 and 1 reflecting its certainty in the decision. The original and modified criteria, as well as the structured prompt can be viewed in the supplementary materials.

**Pipeline**

We developed an integrated screening pipeline in Python and deployed it through a Streamlit interface [32]. The interface accepts CSV files containing titles and abstracts, standardizes input fields, and allows users to specify inclusion and exclusion criteria. An optional training phase lets reviewers label a small sample of abstracts, and this feedback is saved for use in later runs. While we did not test this feature in the current paper, it is intended to be helpful in settings where ground-truth inclusion decisions have not yet been established, allowing early human judgments to guide model behavior and support calibration before full screening is completed.

OLIVER supports two complementary screening modes. First, a real-time screening mode allows abstracts to be processed sequentially within the application interface, which is useful for small-scale testing and criterion iteration. Second, a parallel processing mode, which is recommended for full reviews, enables large batches of abstracts to be screened simultaneously using a JSONL-based request format and custom parallel processing scripts. These scripts are configurable with respect to rate limits and token budgets, allowing screening speed to be optimized based on the constraints of the selected model provider. To reduce data loss during long runs, intermediate results are saved automatically and can be resumed if interrupted. Beyond screening, OLIVER provides built-in evaluation and analysis tools to compare model outputs against human screening decisions. These include confusion matrices, sensitivity, specificity, overall accuracy, and receiver operating characteristic area under the curve (ROC-AUC). Importantly, OLIVER also enables calibration analysis through model-reported confidence outputs, including the computation of Brier scores and expected calibration error (ECE), allowing assessment of how well stated confidence aligns with actual screening correctness.

Once the inclusion and exclusion criteria have been inputted on the frontend, the system then generates a JSONL (JSON Lines) request file in which each abstract is automatically paired with a unified prompt. This prompt embeds the abstract, lists the criteria explicitly, and applies a strict checklist that forces stepwise "reasoning" and fixed output formatting. Each abstract can also be screened multiple times to assess stability and replicate-level agreement, but this feature was not explored in the present paper.

For large-scale inference, we built custom parallel-processing scripts for Claude, Gemini, Grok, and OpenAI models. These were adapted from the OpenAI parallel-processing script [33] but rewritten to align with the rate limits, batching constraints, and endpoint formats of each API. The scripts read the JSONL request file, send requests in parallel with automatic retries, and save model outputs to structured JSONL files. The pipeline then converts these outputs to CSV, extracts model decisions and confidence scores, and aggregates results across replicates to compute agreement statistics.

All of this is wrapped into a Streamlit user interface, which guides the user through each step through simple clicks, allowing users with no coding background to upload datasets, run models, view summaries, and download results. This setup gives us a reproducible and high-throughput framework for benchmarking LLM-based abstract screening across multiple models and providers.

## Analysis

We assessed performance by comparing AI decisions with human reviewer decisions at two levels. First, we evaluated agreement at the level of studies that passed abstract screening only and were sent to full-text screening (Table 4). We then compared screening results to final included records.

The titles marked as "include" by human reviewers were treated as the target set, and the AI's includes were compared directly against this set. From this, we calculated recall (the proportion of human includes detected by the AI), precision (the proportion of AI includes that matched human includes), and the size of the overlap between the two sets.

Second, we examined performance record by record. We combined human include and human exclude files to assign each abstract a binary ground-truth label, then compared these labels with the LLMs predicted include or exclude decision. This allowed us to build a confusion matrix containing true positives, false positives, false negatives, and true negatives. Using these counts, we calculated sensitivity, specificity, overall accuracy, and the area under the ROC curve (AUC). We also computed Brier scores and ECE to quantify how closely the model's confidence values aligned with correct predictions. Together, these metrics describe how often the AI correctly identifies studies that should be kept in the review, how often it rejects irrelevant records, the number of false positive includes, and how reliable its decision confidence is relative to human judgment.

# Results

## Single model performance

**Review 1:** Review 1 consisted of 821 abstracts. 191 were sent to full-text screening out of which 63 were included in the final paper [25]. Performance across models varied widely (Tables 1 & 3).

At the full-text screening stage, models differed in abstract selectivity (Table 1). Sensitivity ranged from 31.41% to 80.63%, with gpt-5-mini forwarding the largest proportion of abstracts to full-text review (154/191; 80.63%) at the cost of low precision (47.83%), indicating substantial over-inclusion. In contrast, gpt-5 and grok-4-fast exhibited more selective screening behaviour, combining moderate sensitivity (57.07% and 58.64%) with high specificity (>93%) and strong precision (>72%). Claude Sonnet followed a consistently conservative pattern, advancing fewer abstracts (31.41% sensitivity) while achieving the highest specificity (97.11%).

These full-text screening behaviours largely carried through to final inclusion decisions. Several models captured most studies included for final analysis with some achieving perfect recall (Table 3). gpt-5-Mini achieved the highest TP count with 100% sensitivity, followed closely by gpt-5-nano & gemini flash (62/63; 98.41%). gpt-5 achieved strong overall balance with 57 true positives (90.48%) and high specificity (88.56%). grok-4-fast performed similarly well, identifying 58 true positives (92%) with high specificity (87.2%), making it the most stable lightweight model.

False positive counts varied substantially across models in Review 1 (Tables 1 & 3). At the full-text stage, high-sensitivity models such as gpt-5-nano (168 false positives), Claude Haiku (169), and Gemini Flash (95) generated the largest number of false positives, while more conservative models including Claude Sonnet (18), gpt-5 (30), and grok-4-fast (42) produced substantially fewer. This pattern persisted at the final inclusion stage, where models with strong recall again incurred higher false positive counts, including Gemini Flash (173) and gpt-5-nano (260). In contrast, Claude Sonnet (51) and gpt-5 (82) remained the most selective.

Overall, false positive burden increased sharply as models prioritized sensitivity, often differing in their trade-offs. Models like Gemini Flash and gpt-5-mini showed high sensitivity but modest specificity, producing a substantial false positive rate. In contrast, Claude Sonnet demonstrated high specificity at the final include stage (94%) but captured only 33 of 63 true positives (52% sensitivity), reflecting a more conservative screening pattern. AUC values ranged between 0.55 and 0.8 across models (Figures 1 & 2), with gpt-5 providing the strongest discrimination.

Calibration across models was poor despite strong accuracy at both the full-text and final inclusion stages (Tables 1 & 3). At full-text screening, Brier scores (0.13–0.48) and ECE values (0.05–0.51) indicated substantial misalignment between predicted confidence and observed outcomes (Table 1). gpt-5 and Claude Sonnet showed the strongest calibration, with low Brier scores and minimal ECE, while higher-sensitivity models such as gpt-5-mini, Gemini Flash, and Claude Haiku were poorly calibrated despite reasonable discrimination. This pattern persisted for final includes (Table 3), where models with strong recall exhibited some of the highest Brier scores and ECE suggesting that high sensitivity did not translate to reliable confidence estimates.

**Review 2:** Review 2 consisted of 7,741 abstracts. 659 were sent to full-text screening, and 71 for final analysis (Table 2). True positive counts were lower across models compared to Review 1 (Tables 2 & 4).

At the full-text screening stage, models showed pronounced differences in selectivity and sensitivity (Table 2). Without prompt modification, most models were highly conservative, with sensitivities below 30% but very high specificity. gpt-5-mini identified the largest number of full-text includes under the original criteria (294/659; 44.61% sensitivity) while maintaining high specificity (94.20%). Prompt modification substantially altered behaviour for some models: grok-4-fast with modified criteria advanced 408 abstracts to full-text (61.91% sensitivity), but at the cost of reduced specificity (77.83%), while Gemini Flash saw sensitivity increase from 6.83% to 29.44% with a corresponding drop in specificity.

At the final-include screening stage (Table 4), the best-performing model was grok-4-fast with modified criteria (Appendix), which recovered 60 of 71 true positives (84.51% sensitivity). gpt-5-mini also showed strong sensitivity with 57 true positives (80.28% sensitivity) while maintaining high specificity (91.51%). gpt-5 provided a more conservative profile, capturing 37 true positives (52.11%) but with near-perfect specificity (97.51%).

Prompt modifications had mixed effects. For Gemini Flash, the modified prompt improved sensitivity from 21.13% to 54.93%, although precision remained low. For gpt-5-nano, the modified prompt reduced sensitivity (43.66% to 36.62%) but raised AUC from 0.70 to 0.79 (Figures 3 & 5). Overall, no model reached near-complete capture of the human includes, but several recovered more than half of the eligible studies, and prompt design played a meaningful role in shaping performance.

False positive counts were substantially higher in Review 2 and were strongly influenced by prompt modification (Tables 2 & 4). Under the original, unmodified criteria, models generally produced low false positive counts, with gpt-5 (87), grok-4-fast (91), and Gemini Flash (19) remaining highly selective at the full-text stage. Applying the stratified, inclusion-biased prompt substantially increased false positives for several models, most notably grok-4-fast (1,548 false positives) and Gemini Flash (374), reflecting a shift toward higher sensitivity. This pattern persisted at the final inclusion stage, where prompt-modified grok-4-fast (1,896) and prompt-modified Gemini Flash (529) generated the largest false positive burdens.

Similar to Review 1, calibration across models in Review 2 was poor despite high accuracy (Tables 2 & 4). At both full-text and final inclusion stages, Brier scores and ECE values were consistently elevated, particularly for high-sensitivity configurations such as prompt-modified grok-4-fast and Gemini Flash. More conservative models like gpt-5 showed comparatively better calibration, but even these exhibited substantial misalignment between confidence estimates and observed outcomes (Tables 2 & 4). Overall, models that maximized recall in this larger screening task produced the least reliable confidence estimates.

**Table 1: Individual Model Performance for Review 1 - Full-Text Includes**

| Review | Model | True Positives | False Positives | Sensitivity (%) | Specificity (%) | Accuracy (%) | Precision (%) | AUC | Brier Score | ECE |
|---|---|---|---|---|---|---|---|---|---|---|
| Review 1 (821 abstracts) | gpt-4o-mini | 106/191 | 88 | 55.5 | 85.8 | 78.67 | 54.64 | 0.51 | 0.37 | 0.51 |
| | gpt-5 | 109/191 | 30 | 57.07 | 94.94 | 85.71 | 78.42 | 0.78 | 0.13 | 0.08 |
| | gpt-5-mini | 140/191 | 82 | 73.3 | 86.73 | 83.56 | 63.06 | 0.75 | 0.22 | 0.19 |
| | gpt-5-nano | 154/191 | 168 | 80.63 | 72.95 | 74.75 | 47.83 | 0.73 | 0.28 | 0.25 |
| | Claude sonnet | 60/191 | 18 | 31.41 | 97.11 | 81.7 | 76.92 | 0.66 | 0.16 | 0.05 |
| | Claude haiku | 130/191 | 169 | 68.06 | 72.02 | 71.07 | 43.48 | 0.61 | 0.48 | 0.45 |
| | Gemini flash | 140/191 | 95 | 73.3 | 84.45 | 81.8 | 59.57 | 0.69 | 0.35 | 0.28 |
| | grok-4-fast | 112/191 | 42 | 58.64 | 93.26 | 85.14 | 72.73 | 0.76 | 0.22 | 0.18 |

**Table 2: Individual Model Performance for Review 2 - Full-Text Includes**

| Review | Model | True Positives | False Positives | Sensitivity (%) | Specificity (%) | Accuracy (%) | Precision (%) | AUC | Brier Score | ECE |
|---|---|---|---|---|---|---|---|---|---|---|
| Review 2 (7741 abstracts) | gpt-4o-mini | 174/659 | 1213 | 26.4 | 82.62 | 77.77 | 12.55 | 0.59 | 0.59 | 0.68 |
| | gpt-5 | 138/659 | 87 | 20.94 | 98.75 | 92.04 | 20.94 | 0.7 | 0.21 | 0.06 |
| | gpt-5-mini | 294/659 | 405 | 44.61 | 94.2 | 89.92 | 42.06 | 0.65 | 0.26 | 0.16 |
| | gpt-5-nano | 97/659 | 267 | 14.72 | 96.17 | 89.13 | 26.65 | 0.64 | 0.3 | 0.35 |
| | Gpt-5-nano* | 127/659 | 358 | 19.27 | 94.82 | 88.24 | 26.19 | 0.67 | 0.31 | 0.36 |

| | Claude haiku | 89/659 | 126 | 13.51 | 96.9 | 75.25 | 41.4 | 0.61 | 0.36 | 0.35 |
|---|---|---|---|---|---|---|---|---|---|---|
| | Gemini flash | 45/659 | 19 | 6.83 | 99.73 | 91.66 | 70.31 | 0.59 | 0.25 | 0.15 |
| | Gemini flash* | 194/659 | 374 | 29.44 | 95.57 | 88.89 | 34.15 | 0.53 | 0.58 | 0.38 |
| | grok-4-fast | 99/659 | 91 | 15.02 | 98.7 | 91.48 | 52.11 | 0.72 | 0.27 | 0.23 |
| | grok-4-fast* | 408/659 | 1548 | 61.91 | 77.83 | 76.45 | 20.86 | 0.57 | 0.66 | 0.62 |

*Note.* * refers to prompt-modified trials

**Table 3: Individual Model Performance for Review 1 - Final Includes**

| Review | Model | True Positives | False Positives | Sensitivity (%) | Specificity (%) | Accuracy (%) | Precision (%) | AUC | Brier Score | ECE |
|---|---|---|---|---|---|---|---|---|---|---|
| Review 1 (821 abstracts) | gpt-4o-mini | 45/63 | 149 | 71.43 | 80.05 | 79.38 | 23.2 | 0.55 | 0.62 | 0.67 |
| | gpt-5 | 57/63 | 82 | 90.48 | 88.56 | 88.72 | 41.01 | 0.8 | 0.36 | 0.4 |
| | gpt-5-mini | 63/63 | 159 | 100 | 78.66 | 80.32 | 28.38 | 0.76 | 0.46 | 0.53 |
| | gpt-5-nano | 62/63 | 260 | 98.41 | 65.24 | 67.82 | 19.25 | 0.75 | 0.42 | 0.53 |
| | Claude sonnet | 33/63 | 45 | 52.38 | 94 | 90.77 | 42.31 | 0.68 | 0.37 | 0.38 |
| | Claude haiku | 54/63 | 245 | 85.71 | 66.44 | 67.97 | 18.06 | 0.61 | 0.69 | 0.69 |
| | Gemini flash | 62/63 | 173 | 98.41 | 76.56 | 78.28 | 26.38 | 0.75 | 0.63 | 0.51 |
| | grok-4-fast | 58/63 | 96 | 92.06 | 87.2 | 87.58 | 37.66 | 0.76 | 0.53 | 0.4 |

**Table 4: Individual Model Performance for Review 2 - Final Includes**

| Review | Model | True Positives | False Positives | Sensitivity (%) | Specificity (%) | Accuracy (%) | Precision (%) | AUC | Brier Score | ECE |
|---|---|---|---|---|---|---|---|---|---|---|
| | **gpt-4o-mini** | 33/71 | 1354 | 46.48 | 82.1 | 81.77 | 2.38 | 0.66 | 0.64 | 0.79 |
| | **gpt-5** | 37/71 | 188 | 52.11 | 97.51 | 97.09 | 16.44 | 0.7 | 0.31 | 0.43 |
| | **gpt-5-mini** | 57/71 | 642 | 80.28 | 91.51 | 91.41 | 8.15 | 0.72 | 0.33 | 0.48 |
| | **gpt-5-nano** | 10/71 | 152 | 14.08 | 96.62 | 95.34 | 6.17 | 0.69 | 0.36 | 0.55 |
| | **Gpt-5-nano*** | 26/71 | 459 | 36.62 | 93.88 | 93.34 | 5.36 | 0.79 | 0.3 | 0.57 |
| | **Claude haiku** | 24/71 | 191 | 33.8 | 95.89 | 94.95 | 11.16 | 0.56 | 0.6 | 0.7 |
| | **Gemini flash** | 15/71 | 49 | 21.13 | 99.35 | 98.62 | 23.44 | 0.53 | 0.69 | 0.47 |
| | **Gemini flash*** | 39/71 | 529 | 54.93 | 92.93 | 92.57 | 6.87 | 0.59 | 0.82 | 0.57 |
| Review 2 (7741 abstracts) | **grok-4-fast** | 31/71 | 159 | 43.66 | 97.9 | 97.39 | 16.32 | 0.7 | 0.45 | 0.58 |
| | **grok-4-fast*** | 60/71 | 1896 | 84.51 | 74.94 | 75.03 | 3.07 | 0.73 | 0.8 | 0.76 |

*Note.* * refers to prompt-modified trials

**Figure 1: AUC's OpenAI Models - Review 1 - Final Includes**

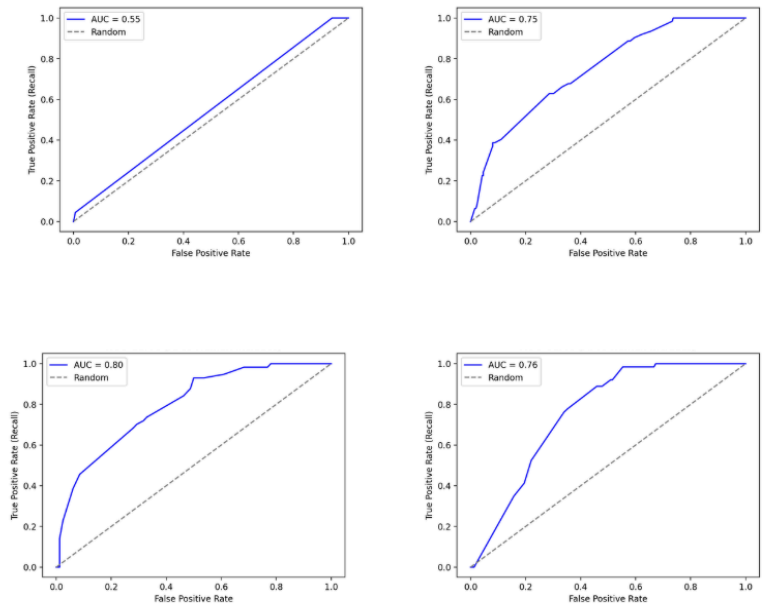

*Note.* From left to right, gpt-4o-mini (0.55), gpt-5-nano (0.75), gpt-5 (0.8), gpt-5-mini (0.76).

**Figure 2: AUC's Other Models, Review 1 - Final Includes**

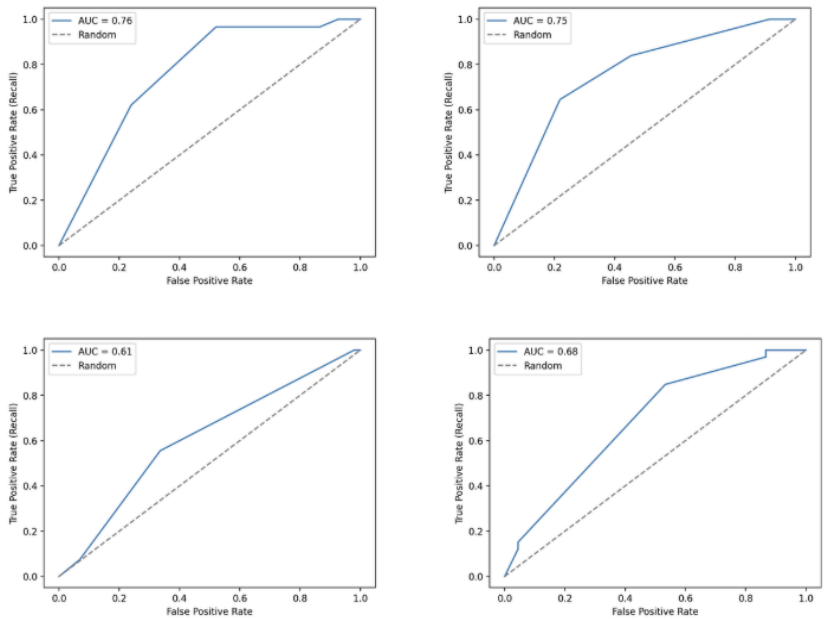

*Note.* From left to right, grok-4-fast (0.76), gemini-flash (0.75), claude haiku (0.61), claude sonnet (0.68).

**Figure 3: AUC's OpenAI Models, Review 2 - Final Includes**

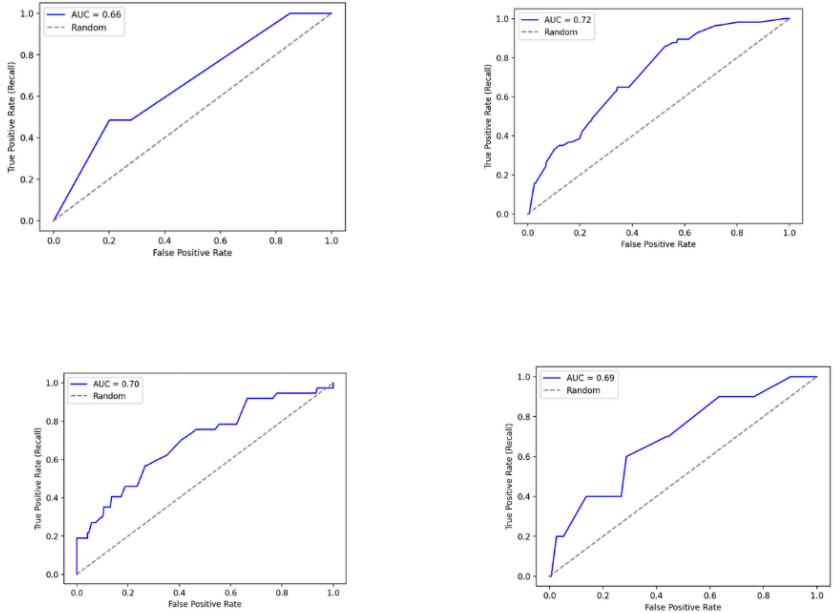

*Note.* From left to right, gpt-4o-mini (0.66), gpt-5-mini (0.72), gpt-5 (0.7), gpt-5-nano (0.69)

**Figure 4: AUC's Other Models, Review 2 - Final Includes**

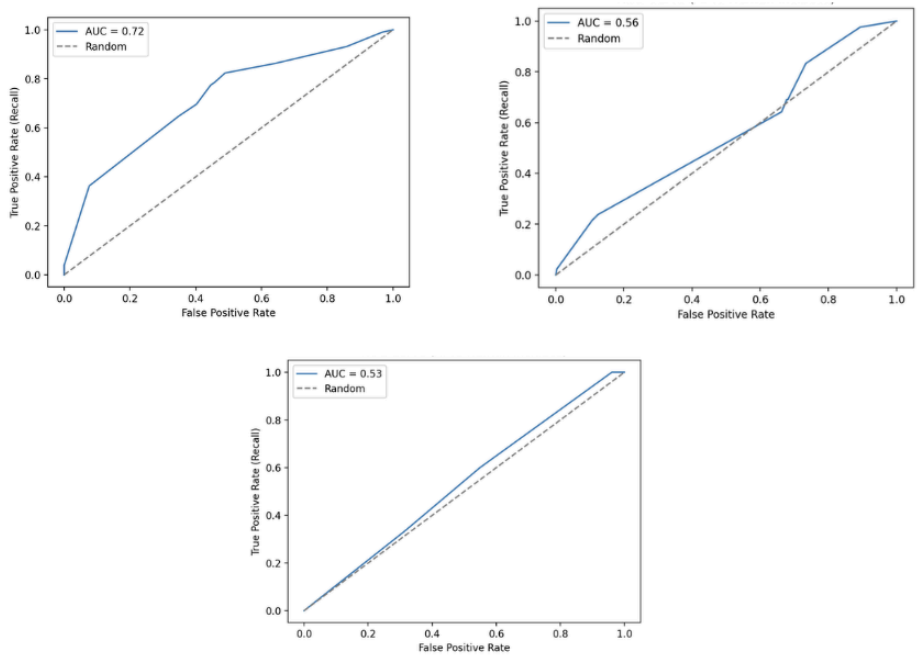

*Note.* From left to right, grok-4-fast (0.72), claude haiku (0.56), gemini-flash (0.53)

**Figure 5: AUC's Modified Prompt, Review 2 - Final Includes**

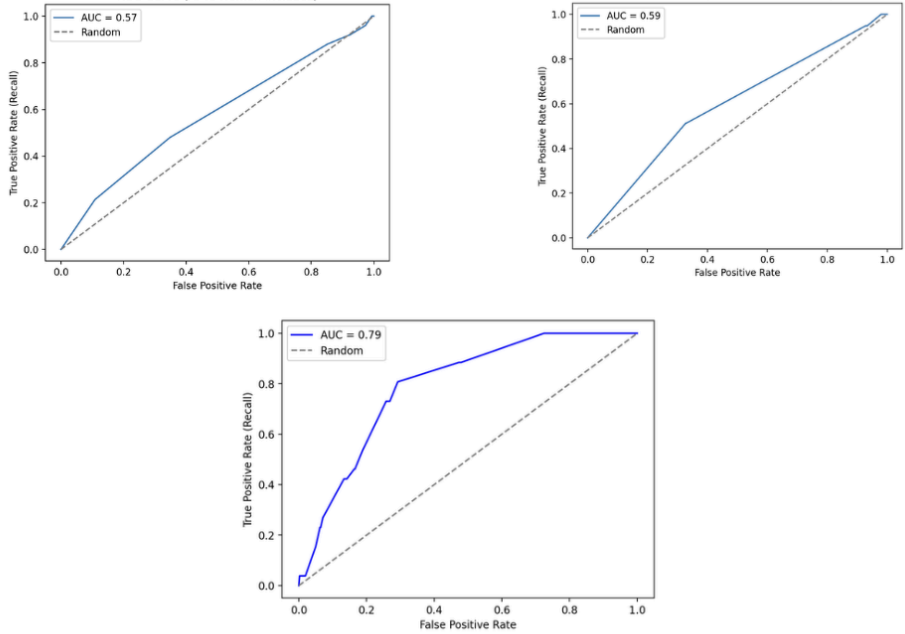

*Note.* From left to right, grok-4-fast (0.57), gemini-flash (0.59), & gpt-5-nano (0.79)

## Actor-Critic Performance

We then evaluated whether combining models in an actor-critic framework improved screening behaviour. Based on individual model performance, grok-4-fast was selected as the actor and gpt-5-nano as the critic due to their balance of screening performance, calibration, and computational efficiency (Tables 1–3). Additionally, both models are lightweight and low-latency, enabling higher throughput and more flexible rate limiting, which is critical for scalable double-screening workflows. grok-4-fast demonstrated consistently strong standalone performance across both reviews, with high sensitivity and specificity in Review 1 and very high specificity in Review 2 (Tables 1–2), making it well suited for primary screening. gpt-5-nano exhibited a more conservative decision profile with comparatively lower Brier scores among lightweight models (Table 2), supporting its use as a critic focused on re-evaluation rather than recall maximization. Three configurations were tested: Any Include, Critic Vetoes, and Agreement Required (Tables 5 & 6). The aim was to explore whether pairing models could increase true positive recovery without overwhelming false positive growth. Additionally, confidence was computed as the mean of the actor and critic confidence scores, and this aggregated confidence was used to derive calibration (Brier score, ECE).

We also evaluated a second actor-critic pairing with gpt-5-mini as the actor and grok-4-fast as the critic (Tables 7 and 8). This pairing was designed to test whether a primary model with advanced "reasoning" capabilities such as gpt-5-Mini, which exhibited strong sensitivity in both Reviews (Tables 3 & 4), when combined with the best lightweight, cheaper model with balanced performance across both reviews (grok-4-fast; Tables 3 & 4) as a critic would further improve recall, discrimination, and calibration.

### Review 1

At the full-text screening stage, actor (grok-4-fast) – critic (gpt-5-nano) configurations improved discrimination relative to individual models (Table 5). Both Any Include and Critic Vetoes configurations advanced over 70% of full-text includes (136/191 and 135/191), while maintaining specificity around 86–87% and achieving higher AUCs (0.81–0.82). The Agreement Required configuration was more conservative, forwarding fewer abstracts (63.87% sensitivity) but with slightly higher specificity (87.8%). Overall, pairing models improved balance at the full-text stage without the extreme over-inclusion seen in high-sensitivity single-model configurations.

At the final-include screening stage (Table 6) for the same model pairing, the Any Include configuration achieved the highest sensitivity (100%), recovering all 63 final includes, with modest specificity (79.5%). The Critic Vetoes configuration performed similarly (62 true positives; 98.41% sensitivity; 78.8% specificity), indicating that the critic did little to suppress the actor's decisions for this dataset. The Agreement Required configuration produced a more balanced profile, capturing 60 true positives (95.24% sensitivity) while improving specificity (81.6%), making it the most conservative and stable of the three. Specificity for the any include, and critic-veto conditions were lower than the constituting individual models (Tables 3 & 6), but was highest for the agreement condition (84.9%). The Any Include configuration also yielded the highest AUC (0.92; Figure 6).

Actor–critic configurations showed similar false positive burdens with comparable specificity at the full-text screening stage (Table 5): Any Include (86 FP; 87.0%), Critic Vetoes (81 FP; 86.2%), and Agreement Required (76 FP; 87.8%). At the final inclusion stage, differences became more pronounced (Table 6). Any Include (265 FP; 64.19%) and Critic Vetoes (256 FP; 65.41%) configurations incurred substantial false positives. In contrast, Agreement Required configuration markedly reduced false positives (112 FP) while achieving substantially higher specificity (84.86%), yielding the most balanced actor–critic configuration in Review 1.

Calibration improved substantially under the actor–critic framework (Tables 5 and 6). At full-text screening, all configurations showed low Brier scores (0.14–0.16) and low ECE values (~0.10), representing a marked improvement over individual models. This benefit largely persisted at the final inclusion stage, particularly for the Agreement Required configuration, which achieved the lowest Brier score (0.13) while maintaining high sensitivity.

For Review 1, the gpt-5-mini–grok-4-fast pairing achieved consistently high discrimination across configurations. At the full-text stage, sensitivity ranged from 70.68% to 77.49% (135–148 true positives), with lower false positive counts than the primary pairing, particularly under Agreement Required (59 FP), while maintaining strong AUCs of 0.86 and improved calibration (Brier 0.12–0.13; ECE 0.06–0.08). At the final inclusion stage, Any Include configuration recovered all true positives (100% sensitivity), while Agreement Required and Critic Vetoes options preserved similarly high sensitivity (98.41%) with improved specificity and fewer false positives. Importantly, the prior model pairing of grok-4-fast & gpt-5-nano yielded fewer false positives (Tables 6 & 8) compared to the gpt-5-mini & grok-4-fast pairing. Calibration remained stable, with Brier scores of 0.13–0.16 and AUCs up to 0.96, indicating strong reliability alongside high recall. These results indicate that model pairing produced more reliable confidence estimates alongside competitive screening performance.

**Review 2**

At the full-text screening stage, both actor–critic configurations remained highly selective in Review 2 (Table 5). For the grok-4-fast - gpt-5-nano pairing sensitivity was low across all configurations, ranging from 10.62% to 23.22%, while specificity exceeded 96% and reached 99.54% under Agreement Required. The Any Include configuration identified the most full-text includes (153/659), but gains in recall were modest compared to Review 1, highlighting the increased difficulty of this larger screening task. Overall, actor–critic pairing did not substantially improve full-text recall in Review 2 but preserved strong exclusion performance.

Similarly, actor–critic performance was constrained at the final-include screening stage for Review 2 (Table 6). The best-performing configuration was Any Include, which recovered 31 of 71 true positives (43.66% sensitivity). Both Critic Vetoes (25 true positives; 35.21% sensitivity) and Agreement Required (22 true positives; 30.99% sensitivity) exhibited lower sensitivity. Across all configurations, specificity remained high, and accuracy exceeded 94%, reflecting strong agreement with human excludes but limited power to detect true includes. AUC's were also high ranging from 0.83-0.84 across configurations (Figure 7).

In Review 2, actor–critic configurations differed sharply in false positive control and specificity (Tables 5 & 6). At the full-text stage, Agreement Required generated very few false positives (32 FP; 99.54%), compared with Critic Vetoes (259 FP; 96.29%) and Any Include (259 FP; 96.08%). This pattern persisted at the final inclusion stage, where Agreement Required again minimized false positives (80 FP; 98.94%), while Any Include (396 FP; 94.77%) and Critic Vetoes (367 FP; 95.15%) produced substantially higher false positive counts. Although Agreement Required reduced sensitivity, it consistently offered the strongest specificity and lowest screening burden in this large, low-prevalence review.

Despite limited improvements in recall, calibration improved markedly under the actor–critic framework for Review 2 (Tables 5 and 6). Brier scores at the full-text stage were below 0.10 across all configurations, with ECE values as low as 0.05, substantially outperforming single-model calibration. This pattern persisted at the final inclusion stage, where all configurations maintained low Brier scores despite modest sensitivity.

In Review 2, the gpt-5-mini–grok-4-fast actor–critic pairing showed modest recall gains relative to the primary pairing but preserved the same trade-offs. At the full-text stage, Any Include achieved the highest sensitivity (23.98%; 158/659) with a high false positive burden (604 FP), while Critic Vetoes reduced false positives (517 FP) at lower sensitivity (18.66%), and Agreement Required minimized false positives (303 FP) but with low sensitivity (9.41%). AUC remained stable at 0.70, with improved calibration under stricter configurations. At the final inclusion stage, Any Include recovered 39 of 71 true positives (54.93%) but generated substantial false positives (723 FP), whereas Agreement Required again offered the strongest specificity (95.82%) and lowest false positive count (350 FP) at reduced sensitivity (21.13%).

**Table 5: Actor (grok-4-fast) - Critic (gpt-5-nano) Model: Full-Text Includes**

| Review | Configuration | True Positives | False Positives | Sensitivity (%) | Specificity (%) | Accuracy (%) | Precision (%) | AUC | Brier Score | ECE |
|---|---|---|---|---|---|---|---|---|---|---|
| Review 1 (821 abstracts) | Any Include | 136/191 | 86 | 71.2 | 87 | 83.29 | 62.67 | 0.82 | 0.14 | 0.1 |
| | Critic Vetoes | 135/191 | 81 | 70.68 | 86.2 | 82.56 | 61.1 | 0.81 | 0.16 | 0.12 |

| | Agreement Required | 122/191 | 76 | 63.87 | 87.8 | 82.19 | 61.62 | 0.8 | 0.16 | 0.11 |
| Review 2 (7741 abstracts) | Any include | 153/659 | 274 | 23.22 | 96.08 | 89.79 | 35.83 | 0.72 | 0.09 | 0.08 |
| | Critic Vetoes | 133/659 | 259 | 20.18 | 96.29 | 89.73 | 33.93 | 0.72 | 0.09 | 0.08 |
| | Agreement Required | 70/659 | 32 | 10.62 | 99.54 | 91.87 | 68.63 | 0.72 | 0.08 | 0.05 |

Table 6: Actor (grok-4-fast) - Critic (gpt-5-nano) Model: Final Includes

| Review | Configuration | True Positives | False Positives | Sensitivity (%) | Specificity (%) | Accuracy (%) | Precision (%) | AUC | Brier Score | ECE |
|---|---|---|---|---|---|---|---|---|---|---|
| Review 1 (821 abstracts) | Any Include | 63/63 | 154 | 100 | 79.47 | 81.06 | 29.03 | 0.91 | 0.15 | 0.23 |
| | Critic Vetoes | 62/63 | 159 | 98.41 | 78.8 | 80.32 | 28.05 | 0.84 | 0.19 | 0.27 |
| | Agreement Required | 60/63 | 138 | 95.24 | 81.6 | 82.66 | 30.3 | 0.84 | 0.16 | 0.25 |
| Review 2 (7741 abstracts) | Any include | 31/71 | 396 | 43.66 | 94.77 | 94.29 | 7.26 | 0.84 | 0.05 | 0.15 |
| | Critic Vetoes | 25/71 | 367 | 35.21 | 95.15 | 94.59 | 6.38 | 0.83 | 0.05 | 0.14 |
| | Agreement Required | 22/71 | 80 | 30.99 | 98.94 | 98.31 | 21.57 | 0.83 | 0.03 | 0.19 |

**Table 7: Actor (gpt-5-mini) - Critic (grok-4-fast) Model: Full-Text Includes**

| Review | Configuration | True Positives | False Positives | Sensitivity (%) | Specificity (%) | Accuracy (%) | Precision (%) | AUC | Brier Score | ECE |
|---|---|---|---|---|---|---|---|---|---|---|
| Review 1 (821 abstracts) | Any Include | 148/191 | 89 | 77.49 | 85.71 | 83.78 | 62.45 | 0.86 | 0.13 | 0.08 |
| | Critic Vetoes | 139/191 | 68 | 72.77 | 89.09 | 85.26 | 67.15 | 0.86 | 0.12 | 0.07 |
| | Agreement Required | 135/191 | 59 | 70.68 | 90.53 | 85.87 | 69.59 | 0.86 | 0.12 | 0.06 |
| Review 2 (7741 abstracts) | Any include | 158/659 | 604 | 23.98 | 92.25 | 86.92 | 20.73 | 0.7 | 0.11 | 0.12 |
| | Critic Vetoes | 123/659 | 517 | 18.66 | 93.36 | 87.54 | 19.22 | 0.7 | 0.1 | 0.12 |
| | Agreement Required | 62/659 | 303 | 9.41 | 96.11 | 89.35 | 9.41 | 0.7 | 0.09 | 0.1 |

**Table 8: Actor (gpt-5-mini) - Critic (grok-4-fast) Model: Final Includes**

| Review | Configuration | True Positives | False Positives | Sensitivity (%) | Specificity (%) | Accuracy (%) | Precision (%) | AUC | Brier Score | ECE |
|---|---|---|---|---|---|---|---|---|---|---|
| Review 1 (821 abstracts) | Any Include | 63/63 | 174 | 100 | 76.8 | 78.6 | 26.58 | 0.95 | 0.16 | 0.23 |
| | Critic Vetoes | 62/63 | 145 | 98.41 | 80.67 | 82.04 | 29.95 | 0.96 | 0.14 | 0.21 |
| | Agreement Required | 62/63 | 132 | 98.41 | 82.4 | 83.64 | 31.96 | 0.96 | 0.13 | 0.2 |
| Review 2 (7741 abstracts) | Any include | 39/71 | 723 | 54.93 | 91.37 | 91.06 | 5.12 | 0.84 | 0.08 | 0.19 |
| | Critic Vetoes | 28/71 | 640 | 39.44 | 92.69 | 92.25 | 4.38 | 0.84 | 0.07 | 0.19 |

|  | Agreement Required | 15/71 | 350 | 21.13 | 95.82 | 95.19 | 4.11 | 0.8 | 0.08 | 0.17 |

**Figure 6: AUC's Actor (grok-4-fast) - Critic (gpt-5-nano) Configuration, Review 1 - Final Includes**

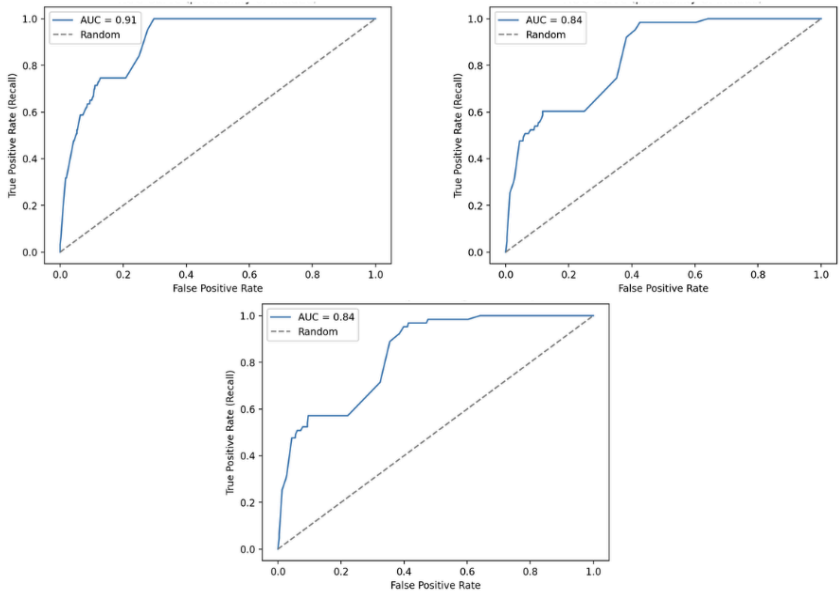

*Note.* From left to right, any include (0.91), critic has veto (0.84), require agreement configuration (0.84).

**Figure 7: AUC's Actor (grok-4-fast) - Critic (gpt-5-nano), Review 2 - Final Includes**

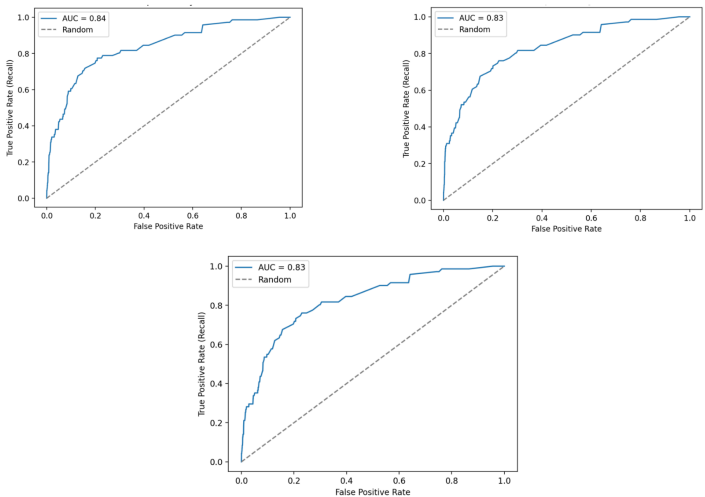

*Note.* From left to right, any include (0.84), require agreement (0.83), critic has veto configuration (0.83).

**Figure 8: AUC's Actor (gpt-5-mini) - Critic (grok-4-fast), Review 1 - Final Includes**

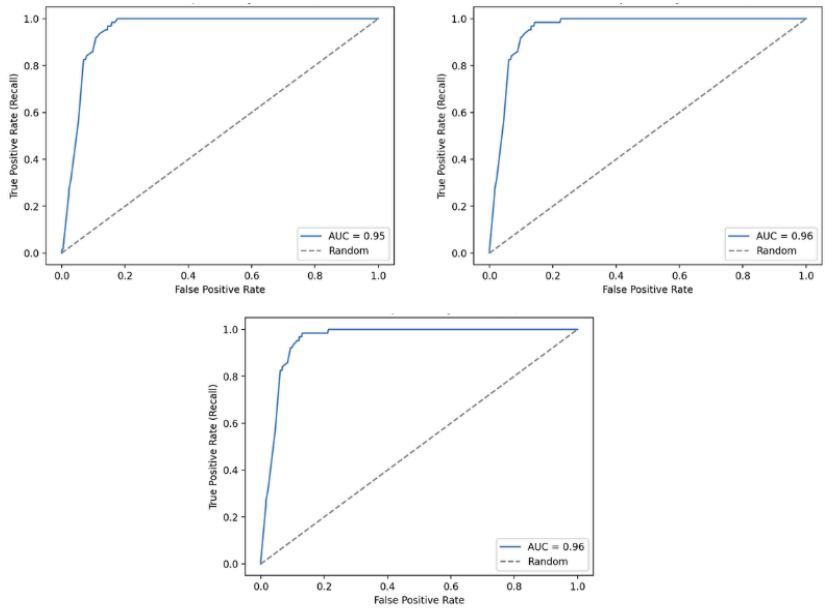

*Note.* From left to right, any include (0.95), require agreement (0.96), critic has veto configuration (0.96).

**Figure 9: AUC's Actor (gpt-5-mini) - Critic (grok-4-fast), Review 2 - Final Includes**

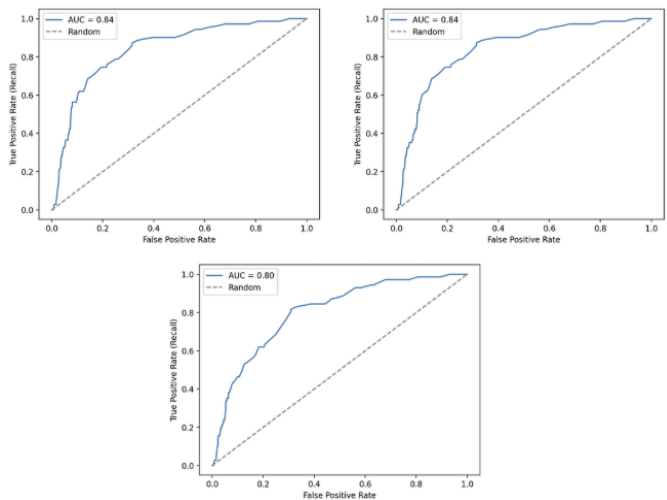

*Note.* From left to right, any include (0.84), critic has veto (0.84), agreement required configuration (0.8)

**Investigating Review 2 Performance**

Given OLIVER's low recall on Review 2, we investigated whether the characteristics of the final includes such as publication date, abstract content or open-access status may have hampered performance. Out of 71 final includes for Review 2, 22 were published prior to 2000, 25 between 2000-2009, 22 between 2010-2020, and 2 after 2020. In the actor (gpt-5-mini) - critic (grok) pipeline with the any-include configuration for Review 2, OLIVER missed 10 (31%) pre-2000 articles, 15 (47%) 2000-2009 articles, and 7 (21.88%) 2010-2020 articles. On the contrary, out of the 63 final includes for Review 1, 16 (25%) were published between 2010-2020, with 47 published after 2020 (75%), which may explain the perfect recall.

Moreover, abstract length (in characters) varied comparably by time-period published for Review 2, with articles published pre-2000 having average abstract lengths of 1072.4, articles published from 2000-2010 with 1554.9, articles published from 2010-2020 with 1619.31 and articles post 2020 with 1771 which may have affected performance. The percentage of papers missing abstracts were similar across both reviews, with Review 1 missing 25 (3%) and Review 2 missing 224 (2.9%). Median abstract lengths differed as well and for Review 1 was 1779 characters, which was higher than the median abstract length for Review 2 at 1570 characters. This translates to approximately 30-40 fewer words for the LLM to go through during screening. A Mann-Whitney Test (Appendix) comparing these features across the full dataset and final includes for both reviews found significant differences in abstract length (in characters; $p<.001$), and publication year ($p<.001$).

Finally, we used the Unpaywall [34] module and Directory of Open Access Journals [35] to investigate open-access status for both reviews. We expected that reviews with a higher proportion of open-access papers would improve LLM performance as they were likely to be better represented in pretraining data. For Review 2, out of 71 final includes, 55 were closed access, and 16 were open access. The any-include configuration of the actor (gpt-5-mini)-critic (grok) which yielded 39 true positives (Table 8), missed 30/55 (55%) of closed-access articles compared to only 2/16 (12%) open-access articles for Review 2. For Review 1, out of 63 final includes, our methodology revealed that 51 (81%) were open-access, and 12 (19%) were closed-access. A manual search of the 12 closed-access included papers for Review 1 revealed that 8 were available as open-access full-texts through alternative sources, bringing the number of open-access articles to 59/63 (93%). Comparing the full datasets for both reviews, nearly two-thirds of all papers in Review 2 (Appendix) were closed-access, whereas for Review 1, 70% of papers were open access (though the full texts for review 2 were not systematically checked for open access status given their older age).

A Fisher's exact test (Appendix) revealed significant differences in open-access status between the two reviews across the full dataset and final include stages ($p<0.001$). However, a key limitation of this analysis to assess open-access status was that papers counted as 'closed-access' papers may have had open-access preprints or manuscripts via Sci-Hub that were missed by our methodology and it should thus be interpreted with caution.

## Discussion

In this paper, we developed and tested an open-source pipeline, OLIVER, against two systematic reviews to measure its efficacy in abstract screening at scale. We first explored single-model-screening, and then tested an actor-critic configuration.

**Overall Performance Across Reviews**

For Review 1, OLIVER performed well (Table 1). The individual models identified an average of 86% of the final includes from abstracts alone. gpt-5-mini performed best (100% recall) followed by Gemini Flash and gpt-5-nano, which both captured 62/63 final includes (98.41%). This level of sensitivity came with reduced specificity, reflected in an overall accuracy of 80% and an average AUC of 0.71. Earlier models such as gpt-4o-mini performed poorly, which was expected due to poorer "reasoning" and 'chain-of-thought' capabilities [36]. Claude Sonnet [28], despite being one of the highest-end "reasoning" models, also struggled to detect true positives, possibly because it applied the criteria in an overly rigid or deliberative way. False positive rates in Review 1 were comparatively lower across single models, although high-sensitivity models still generated more false positives than conservative models (Tables 1 & 2).

Performance on Review 2 was mixed (Table 2). Individual models identified only 46% of the final includes from the abstract screening phase, with a lower average AUC across models of 0.65 (Table 4). We suspected that this weaker performance was due to the more vague criteria in Review 2. Prompt modification with lightweight models only led to a modest improvement, recovering 58% of the final includes, and a higher average AUC of 0.7 (Table 4). Importantly, this improved sensitivity led to a significant decrease in specificity, with models like grok-4-fast yielding over 1500 false positives with modified criteria, compared to only 91 with the original criteria. This highlights the tradeoff between standard and inclusion-biased prompts in abstract-screening workflows, where the former yields fairly balanced performance though at the cost of a loss of true positives, while the latter improves sensitivity at the cost of increasing second pass screening workloads.

**Dataset Characteristics Significantly Affect Performance**

Our analysis of abstract characteristics (Appendix) suggest that OLIVER's poorer performance on Review 2 was driven by differences in publication era, abstract length and open access status. These dataset features, which are often driven by review topics and date ranges of interest, can significantly affect LLM performance in abstract screening as older papers are more likely to have less detailed abstracts, and be closed-access. Review 2 contained a substantial proportion of older studies that had shorter abstracts, potentially limiting the presence of explicit eligibility relevant signals. Additionally many missed studies in Review 2 were not open access, which appears to affect performance as the model was less likely to have seen the paper during training. It is therefore possible that the more recent and openly available literature in Review 1 was well represented in the training data of contemporary LLMs, meaning that strong performance may partially reflect prior exposure rather than abstract level reasoning alone. Additionally, it is possible that the larger size of Review 2 increased the likelihood of edge cases, further challenging consistent application of inclusion criteria at the abstract level.

**Calibration & Confidence of LLMs in Abstract Screening**

During screening, we also required that each model provide a confidence output between 0 and 1, which was used to compute Brier scores. For individual model screening, poor calibration was observed based on model-reported confidence across both reviews (Tables 1-4), which aligns with other calibration research [22], [23]. Models that achieved high sensitivity, including those that recovered nearly all final includes, often exhibited some of the poorest calibration, with elevated Brier scores and ECE values, while more conservative models showed comparatively better calibration and lower recall (Tables 1-4). This disconnect replicates prior findings of poor-out-of-the-box calibration exhibited by LLMs [22], [23], indicating that raw confidence outputs from single-model screening is an unreliable indicator of screening correctness. These results indicate that model confidence should not be relied upon during single-model screening, particularly when final decision labels are unknown.

In contrast, the actor–critic framework substantially improved calibration across both reviews, producing consistently lower Brier scores and ECE values even when gains in sensitivity were modest (Tables 5-6). This suggests that pairing models helps stabilize probability estimates and reduce overconfident errors. Furthermore, given that calibration is helpful in establishing human trust in AI-assisted decision making [19], [20], [21], the actor-critic approach may also improve eventual adoption in live abstract screening workflows that lack final decision labels.

**Benefits of the Actor-Critic Approach**

Beyond calibration, the actor critic pipeline, where grok served as the actor and gpt-5-nano as the critic, yielded strong balanced results for both reviews (Tables 5-6). Across all configurations, AUC values were markedly higher than those of individual models, reaching 0.91 for Review 1 and 0.84 for Review 2. While the Any Include and Critic Vetoes yielded similar specificity and false-positive rates to single models, the Agreement Required condition maintained high sensitivity, while also significantly increasing specificity and decreasing the number of false positives (Tables 3 & 6). These improvements, maintained across different model pairings, suggest that the pipeline created a more reliable classification of abstracts with two key factors likely contributing to this pattern.

First, combining two independent model signals reduces random variance in scoring. The actor proposes an initial classification while the critic evaluates that decision, which smooths out inconsistent judgments. Second, the critic provides a corrective influence on overconfident or erratic probabilities, which is reflected in the substantially lower Brier scores across all actor-critic configurations. For example, in Review 1, Brier scores fell from 0.49 for grok and 0.59 for gpt-5-nano in individual settings to as low as 0.13 in the agreement condition (Tables 1 & 6). This improved calibration helps separate true positives and true negatives more cleanly. Another possible contributor to this improved calibration could be the aggregation strategy which included taking the average of model reported confidence scores for the actor-critic pipeline. While this would have slightly reduced variance, it is unlikely to fully account for the magnitude of the observed reductions in Brier scores and ECE's.

Our results suggest that the actor-critic design improves both classification and calibration. Additionally, the actor-critic pipeline with grok-4-fast and gpt-5-nano relies on fast and inexpensive models, allowing large scale screening to be completed quickly and at low cost while maintaining better performance than advanced "reasoning" models. The alternative gpt-5-mini–grok-4-fast pairing achieved comparable discrimination and calibration but at higher computational cost, with limited additional benefit. Overall, the total time required to complete screening for Review 1 was under 5 minutes, with costs ranging from $1 to $3.50 depending on the model. Review 2 took about 1.25 hours to screen due to throttling limits, with costs ranging from $5 to $15 across models. For the actor-critic pipeline, the cost to screen Review 1 was ~$1, and ~$5.5 for Review 2 for the grok-4-fast and gpt-5-nano pairing.

**Lightweight Models Perform Well**

An additional finding from this study is that lightweight models often matched or exceeded more advanced ""reasoning"" models in true positive recovery. Models optimized for speed and efficiency, such as Gemini Flash, gpt-5-mini, and grok-4-fast, consistently captured a larger proportion of final includes than heavier models like Claude Sonnet and gpt-5, albeit with higher false positive rates. In contrast, advanced "reasoning" models tended to exhibit more balanced sensitivity and specificity, but frequently failed to recover a substantial fraction of true includes. These performance differences occurred alongside large cost disparities where lightweight models screened thousands of abstracts at a fraction of the cost and time required by advanced "reasoning" models. This implies that abstract screening may not consistently benefit from deeper deliberative "reasoning", and that fast, low-cost models are better suited to the coarse filtering required at early screening stages.

**Proposed Adoption Strategy**

Importantly, given OLIVER's reduced performance on Review 2, our findings indicate that LLMs are better suited to *augment*, rather than replace, human screening workflows. We therefore propose a hybrid approach in which abstracts are single screened by a human reviewer, with OLIVER, specifically in the actor-critic configuration, serving as secondary screener. Once screening from both sides are complete, decision labels can be compared; we recommend accepting human include labels preferentially over the LLM exclude label for a given record, given concerns about sensitivity and the high cost of missing relevant literature in a systematic review. Additionally, if this approach is leveraged, and single-screening is completed by large teams with more than one human reviewer, we recommend first establishing strong interrater reliability for screening, before using OLIVER, to ensure high quality screening criteria, better alignment, and lower risk for missing true positives. This strategy leverages the relative strengths of LLMs, which in our evaluation were more consistent at excluding irrelevant studies than at identifying true positives. At the same time, this workflow mitigates key limitations observed in our study, including the risk of missing final included studies and difficulties handling edge cases such as records with titles but no or limited content abstracts. Overall, this primary strategy of using OLIVER as a secondary screener offers a pragmatic balance between scalability and oversight, while also providing the additional benefit of triplicate screening through the actor-critic configuration.

**Implications for Human Screening Workflows**

Taken together, our findings show that LLMs can support fast and low cost abstract screening, though their performance depends on dataset complexity, model capabilities, and prompt design. Individual model performance varied widely and was often limited by inconsistent calibration, but the actor critic pipeline reduced these weaknesses by improving classification quality and producing more reliable probability estimates, which led to higher AUC values across both reviews. Although identifying true includes remains challenging for complex datasets with older, less content-heavy abstracts, our results demonstrate that lightweight model combinations can outperform more advanced standalone models while completing screening at scale in minutes to hours through parallel processing and at a fraction of the cost. For example, double-screening of Review 1 & Review 2 (assuming 30 seconds per abstract) would have taken approximately 14 and 129 hours of human labor respectively. On the flipside, our proposed primary adoption strategy would replace the need for double-screening abstracts, requiring only single-screening by review teams. This would reduce human screening time to 7 and 64.5 hours for both reviews respectively, leading to a direct 50% decrease in time spent and labor costs on abstract screening alone.

# Limitations

Several limitations should be considered when interpreting these findings. First, the evaluation was conducted on only two systematic reviews, which limits the generalizability of results across review topics, domains, and evidence types. Furthermore, varying performance in accuracy and sensitivity across both reviews suggest that screening criteria, review topic, and abstract characteristics may significantly affect performance. Second, although multiple widely used

LLMs were tested, the study did not exhaustively evaluate all available models or providers, and performance may differ for other architectures or future model releases. Third, only a limited set of prompting strategies was explored, and more extensive experimentation with alternative prompt structures, "reasoning" constraints, or criteria decompositions may yield different performance trade-offs. Fourth, while OLIVER supports an optional training phase that incorporates human feedback, this feature was not formally evaluated in the current study, and its impact on screening performance and calibration remains an open question. Finally, we initially attempted to require that each LLM provide a brief justification (≤25 words) to examine how screening criteria were being applied. However, constraints related to output length and API throttling prevented this from being implemented at scale. As a result, questions remain about the relationship between calibration, accuracy, and model-generated justifications, and warrants further investigation to clarify whether and how LLMs appropriately apply screening criteria. Future work should address these limitations through broader benchmarking, systematic prompt comparisons, and prospective evaluations in live review settings.

## Conclusion

We present OLIVER, an open-source pipeline for LLM-assisted abstract screening that supports both single-model screening and actor–critic configurations. Across two systematic reviews, single-model screening achieved high sensitivity at low cost, but was frequently limited by poor calibration, high false positive rates, and inconsistent confidence estimates. In contrast, the actor–critic pipeline consistently improved calibration and classification performance, yielding higher AUCs and more reliable probability estimates, particularly in reviews with broader or more ambiguous inclusion criteria. These findings indicate that ensemble configurations, in the form of actor–critic screening, are better suited for AI-assisted abstract-screening workflows. OLIVER demonstrates that structured model pairing can outperform advanced standalone models while enabling rapid, scalable screening, and provides a concrete framework for exploring the integration of LLMs into systematic review workflows with human oversight. Yet, the variability of results and poor overall sensitivity in Review 2 suggests that LLMs do not yet possess the ability to independently complete literature reviews.

# Appendix

## Criteria for Review 1

**Inclusion Criteria:**
**A) Population:** Any study whose target population either has an official mental health diagnosis: Mental health disorders of interest involve:
1- Depression
2- Anxiety
3- Trauma-related disorders
4- Psychotic disorders
5- Personality disorders
6- Bipolar disorders
7- Non-tobacco substance use disorders
8 - Eating disorders
9- Neurodevelopmental disorders (e.g. ASD or ADHD).
OR
Belongs to one of these at-risk populations:
1- Individuals at high risk for psychosis
2- Postpartum individuals
3 - Individuals with trauma-related experiences
4 - Populations considered to be minorities or oppressed populations
**B) Intervention / Exposure:** Studies that utilize digital phenotyping methods to monitor and assess mental health conditions. Studies that measure SDH as part of their data collection or analysis. Examples:
1- Smartphone Apps: like MindLamp allow users to track their symptoms through various surveys, such as the PHQ-9 for mood assessment. The app can also collect passive data and metadata.
2- Wearable devices: Devices such as ActivWatches, Fitbit, or Apple Watch can monitor physiological data, including heart rate, step count, and sleep patterns.
3- Digital Voice Analysis: Voice Recordings: voice analysis tools can assess speech patterns, tone, and rhythm to detect signs of depression, anxiety, or other mental health conditions
**C) Comparator / Context:** No inclusion or exclusion criteria for comparisons or control groups, however, we will look for studies that compare digital phenotyping data across different socio-demographic groups.
**D) Outcome:** There are no strict inclusion or exclusion criteria related to outcomes of interest; however, we are particularly interested in studies that examine the impact of social determinants of health (SDH) on digital phenotyping measures, as well as studies that provide data on the effectiveness or reliability of digital phenotyping tools across different socio-demographic groups.
As long as the study population pertains to mental health—whether through diagnosis, symptom monitoring, or being at risk—and uses digital phenotyping, the specific purpose of the study is less critical. For instance, a study using digital phenotyping to monitor and address obesity in individuals with schizophrenia would still be considered relevant if it meets the other inclusion criteria.
**E) Study Characteristics:** All study designs in which digital phenotyping measures are collected are considered, including randomized controlled trials (RCTs), cohort studies, case-control studies, cross-sectional studies, and qualitative studies.

Focus groups or preparatory work completed prior to data collection are not eligible for inclusion as they do not involve the actual collection of digital phenotyping data but will be used to provide context to our results if they acknowledge SDoH.

Studies must involve primary research rather than systematic reviews, but studies pooling individual patient data across previously collected samples are acceptable. Case reports are acceptable. Theses and preprints are acceptable.

**Exclusion Criteria:**
**A) Population:** Studies not focused on mental health conditions or patients at risk for or being screened or monitored for mental health conditions and symptoms: Excluded mh conditions/disorders:
1- Neurocognitive disorders
2- Solely primary insomnia
3- Nicotine addiction
4- Gambling addiction
5- Nomophobia
AND
Excluded at-risk populations solely based on:
1- Chronic physical health condition or status
2- Studies conducted during or investigating the effects of covid-19 or long covid
3- Primary focus on older populations
4- Individuals experiencing chronic stress
5- Individuals working in high-stress occupations, (nurses, firefighters, etc.)
AND
Studies focused on mental health but targeting caregivers will be excluded.
**B) Intervention/Exposure:** Studies focused on using only ecological momentary assessments or experience sampling methods without any other digital phenotyping technology.
Mobile phone apps that only use self-surveys to collect data.
**C) Comparator/Context:** The only exclusion criterion for setting is studies that are solely based in a laboratory environment.
Study Characteristics: Literatures reviews
The only exclusion criterion for setting is studies that are solely based in a laboratory environment.
Study protocols/focus groups
Abstracts
Opinion-based articles
**D) Other:** Studies focused on using only ecological momentary assessments or experience sampling methods without any other digital phenotyping technology; studies using only these studies would be acceptable if the EMA/ESM is used as a digital phenotyping method - e.g. if response times to EMAs are considered as a data point.

# Criteria for Review 2 (from Benrimoh et al., 2024)

**Inclusion Criteria:**
1. Studies of first episode or subsequent episode psychosis (both affective and non-affective were included, as well as psychosis not otherwise specified) in which the prevalence of prodromal symptoms was established (whether the primary aim of the study or not).

2. Studies of general population cohorts followed prospectively to determine how many people experience a prodrome and eventual psychosis.
3. Studies of populations of patients.
4. Studies that provide the proportions of people who experienced a prodrome (as defined by the study) prior to onset of psychosis.
5. Studies that apply a consistent definition of the prodrome within the study. This definition could range from specific (e.g. meeting a threshold on a specific scale) to general (e.g. a brief description of symptoms), as long as it was consistently applied.

**Exclusion Criteria:**
1. Studies in which experiencing a prodrome was an inclusion criterion, as patients who developed psychosis in these cohorts would, by definition, have had a preceding prodrome, artificially inflating the proportion to 100%.
2. Qualitative studies that did not report prevalence data, as well as protocols, conference proceedings/abstracts, reviews, and case studies/case series.
3. Studies solely of patients with substance-induced psychosis (though studies with a minority of patients with drug-induced psychosis were allowed; it was generally not possible to separate these patients out in prevalence calculations).

## Modified Criteria for Review 2

**Inclusion Criteria:**
**A) Population:**
Studies of first episode psychosis or subsequent episode psychosis (affective, non-affective, or NOS). Studies of general population cohorts followed prospectively to see how many develop a prodrome and eventual psychosis. Here prodrome means period of symptoms contiguous with the onset of psychosis.
**B) Outcome / Exposure:**
Must establish the prevalence of prodromal symptoms (whether or not it is the primary aim).
Must report the proportion of people who had a prodrome prior to onset of psychosis.
Must apply a consistent definition of the prodrome, ranging from specific (e.g., scale threshold) to general (e.g., descriptive symptoms), as long as it is used consistently.
**C) Study Type / Design**
Must be a study of populations (not case series, reviews, or opinion papers).
All quantitative designs are eligible if prevalence is reported.
**D) Setting / Context**
Any geographic or clinical setting is acceptable.
**E) Heuristic**
If the study satisfies B, C, and D, and it analyses a prodrome or the transition into first episode psychosis, lean towards inclusion. When unsure, include as long as the study presents prevalence data related to prodromal symptoms.

**Exclusion Criteria:**
**A) Population**

Studies solely of patients with substance-induced psychosis (minority samples are acceptable if not separated out).
**B) Study Type / Design**
Qualitative studies that do not report prevalence.
Protocols, conference proceedings/abstracts, reviews, and case studies/case series.

## System Prompt

You are assisting with abstract screening.

Inclusion Criteria: {inclusion_criteria}
Exclusion Criteria: {exclusion_criteria}

Title: {title}
Abstract: {abstract}

---

Decide whether to Include or Exclude based on the criteria.

Evaluate in this strict order:
- Inclusion-1 … met? (yes/no)
- Inclusion-2 … met? (yes/no)
- Inclusion-3 … met? (yes/no)
- Exclusion-1 … applies? (yes/no)
- Exclusion-2 … applies? (yes/no)

Rules:
- Include only if ALL inclusions are "yes" AND ALL exclusions are "no".
- If any inclusion is "no" or any exclusion is "yes", Exclude.
- Quote ≤12 words of supporting evidence from the Abstract when you say "yes".

Output exactly two lines:
Decision: YYY or XXX
Confidence: number between 0 and 1

**Mann-Whitney Test Between Review 1 & Review 2**

| Screening Stage | Variable | Median (Review 1) | Median (Review 2) | p-value |
|---|---|---|---|---|
| **Full Dataset** | **Publication Year** | 2021 | 2013 | <0.001 |
| | **Abstract Length (chars)** | 1902 | 1652 | <0.001 |
| **Final Includes** | **Publication Year** | 2022 | 2006 | <0.001 |
| | **Abstract Length (chars)** | 1779 | 1570 | <0.001 |

**Fisher's Exact Test Between Review 1 & Review 2 - Open Access Status**

| Screening Stage | Review | Open-Access | Closed-Access | Odds Ratio | p-value |
|---|---|---|---|---|---|
| **Full Dataset** | **Review 1** | 578 (70%) | 235 (30%) | 4.38 | <0.001 |
| | **Review 2** | 2772 (36%) | 4932 (64%) | | |
| **Final Includes** | **Review 1** | 51 (81%) | 12 (19%) | 15.6 | <0.001 |
| | **Review 2** | 16 (23%) | 55 (77%) | | |

*Note.* The open-access status of some papers for both reviews are unknown due to methodological constraints.

# References

[1] J. R. Polanin, T. D. Pigott, D. L. Espelage, and J. K. Grotpeter, "Best practice guidelines for abstract screening large‑evidence systematic reviews and meta‑analyses," *Res. Synth. Methods*, vol. 10, no. 3, pp. 330–342, Sep. 2019, doi: 10.1002/jrsm.1354.

[2] M. Michelson and K. Reuter, "The significant cost of systematic reviews and meta-analyses: A call for greater involvement of machine learning to assess the promise of clinical trials," *Contemp. Clin. Trials Commun.*, vol. 16, no. 100443, p. 100443, Dec. 2019, doi: 10.1016/j.conctc.2019.100443.

[3] B. C. Wallace, T. A. Trikalinos, J. Lau, C. Brodley, and C. H. Schmid, "Semi-automated screening of biomedical citations for systematic reviews," *BMC Bioinformatics*, vol. 11, no. 1, p. 55, Jan. 2010, doi: 10.1186/1471-2105-11-55.

[4] J. Rathbone, T. Hoffmann, and P. Glasziou, "Faster title and abstract screening? Evaluating Abstrackr, a semi-automated online screening program for systematic reviewers," *Syst. Rev.*, vol. 4, no. 1, p. 80, Jun. 2015, doi: 10.1186/s13643-015-0067-6.

[5] A. Tsertsvadze, Y.-F. Chen, D. Moher, P. Sutcliffe, and N. McCarthy, "How to conduct systematic reviews more expeditiously?," *Syst. Rev.*, vol. 4, no. 1, p. 160, Nov. 2015, doi: 10.1186/s13643-015-0147-7.

[6] J. C. Carver, E. Hassler, E. Hernandes, and N. A. Kraft, "Identifying barriers to the systematic literature review process," in *2013 ACM / IEEE International Symposium on Empirical Software Engineering and Measurement*, IEEE, Oct. 2013, pp. 203–212. doi: 10.1109/esem.2013.28.

[7] J. Rathbone, E. Kaltenthaler, A. Richards, P. Tappenden, A. Bessey, and A. Cantrell, "A systematic review of eculizumab for atypical haemolytic uraemic syndrome (aHUS)," *BMJ Open*, vol. 3, no. 11, p. e003573, Nov. 2013, doi: 10.1136/bmjopen-2013-003573.

[8] A. Gates, C. Johnson, and L. Hartling, "Technology-assisted title and abstract screening for systematic reviews: a retrospective evaluation of the Abstrackr machine learning tool," *Syst. Rev.*, vol. 7, no. 1, p. 45, Mar. 2018, doi: 10.1186/s13643-018-0707-8.

[9] K. E. K. Chai, R. L. J. Lines, D. F. Gucciardi, and L. Ng, "Research Screener: a machine learning tool to semi-automate abstract screening for systematic reviews," *Syst. Rev.*, vol. 10, no. 1, p. 93, Apr. 2021, doi: 10.1186/s13643-021-01635-3.

[10] M. Li, J. Sun, and X. Tan, "Evaluating the effectiveness of large language models in abstract screening: a comparative analysis," *Syst. Rev.*, vol. 13, no. 1, p. 219, Aug. 2024, doi: 10.1186/s13643-024-02609-x.

[11] F. M. Delgado-Chaves *et al.*, "Transforming literature screening: The emerging role of large language models in systematic reviews," *Proc. Natl. Acad. Sci. U. S. A.*, vol. 122, no. 2, p. e2411962122, Jan. 2025, doi: 10.1073/pnas.2411962122.

[12] R. Sanghera *et al.*, "High-performance automated abstract screening with large language model ensembles," *J. Am. Med. Inform. Assoc.*, vol. 32, no. 5, pp. 893–904, May 2025, doi: 10.1093/jamia/ocaf050.

[13] F. Dennstädt, J. Zink, P. M. Putora, J. Hastings, and N. Cihoric, "Title and abstract screening for literature reviews using large language models: an exploratory study in the biomedical domain," *Syst. Rev.*, vol. 13, no. 1, p. 158, Jun. 2024, doi: 10.1186/s13643-024-02575-4.

[14] A. Homiar *et al.*, "Development and evaluation of prompts for a large language model to screen titles and abstracts in a living systematic review," *BMJ Ment. Health*, vol. 28, no. 1, p. e301762, Jul. 2025, doi: 10.1136/bmjment-2025-301762.

[15] C. Cao *et al.*, "Automation of systematic reviews with large language models," *medRxiv*, p. 2025.06.13.25329541, Jun. 13, 2025. doi: 10.1101/2025.06.13.25329541.

[16] C. Cao *et al.*, "Development of prompt templates for large language model-driven screening in systematic reviews," *Ann. Intern. Med.*, vol. 178, no. 3, pp. 389–401, Mar. 2025, doi: 10.7326/ANNALS-24-02189.

[17] M. Z. Andersen, P. Zeinert, J. Rosenberg, and S. Fonnes, "Comparative analysis of Cochrane and non-Cochrane reviews over three decades," *Syst. Rev.*, vol. 13, no. 1, p. 120, May 2024, doi: 10.1186/s13643-024-02531-2.

[18] Z.-Z. Li *et al.*, "From System 1 to System 2: A survey of reasoning Large Language Models," *arXiv [cs.AI]*, Jun. 25, 2025. [Online]. Available: http://arxiv.org/abs/2502.17419

[19] C. Wang, "Calibration in deep learning: A survey of the state-of-the-art," *arXiv [cs.LG]*, Sep. 14, 2025. [Online]. Available: http://arxiv.org/abs/2308.01222

[20] J. Li, Y. Yang, Q. V. Liao, J. Zhang, and Y.-C. Lee, "As confidence aligns: Exploring the effect of AI confidence on human self-confidence in human-AI decision making," *arXiv [cs.HC]*, Jan. 22, 2025. doi: 10.48550/arXiv.2501.12868.

[21] Y. Zhang, Q. V. Liao, and R. K. E. Bellamy, "Effect of confidence and explanation on accuracy and trust calibration in AI-assisted decision making," in *Proceedings of the 2020 Conference on Fairness, Accountability, and Transparency*, New York, NY, USA: ACM, Jan. 2020, pp. 295–305. doi: 10.1145/3351095.3372852.

[22] M. Omar, R. Agbareia, B. S. Glicksberg, G. N. Nadkarni, and E. Klang, "Benchmarking the confidence of large language models in answering clinical questions: Cross-sectional evaluation study," *JMIR Med. Inform.*, vol. 13, no. 1, p. e66917, May 2025, doi: 10.2196/66917.

[23] R. de Oliveira *et al.*, "A study of calibration as a measurement of trustworthiness of large language models in biomedical natural language processing," *JAMIA Open*, vol. 8, no. 4, p. ooaf058, Aug. 2025, doi: 10.1093/jamiaopen/ooaf058.

[24] D. Benrimoh *et al.*, "On the proportion of patients who experience a prodrome prior to psychosis onset: A systematic review and meta-analysis," *Mol. Psychiatry*, vol. 29, no. 5, pp. 1361–1381, May 2024, doi: 10.1038/s41380-024-02415-w.

[25] T. Zamorano *et al.*, "Systematic review: the integration and interpretation of Social Determinants of Health (SDH) into Digital Phenotyping research (DP)," *PsyArXiv*, Dec. 18, 2025. doi: 10.31234/osf.io/bswyf_v1.

[26] W. S. Richardson, M. C. Wilson, J. Nishikawa, and R. S. Hayward, "The well-built clinical question: a key to evidence-based decisions," *ACP J. Club*, vol. 123, no. 3, pp. A12–3, Nov. 1995, doi: 10.7326/acpjc-1995-123-3-a12.

[27] "OpenAI - Pricing," OpenAI Platform. Accessed: Dec. 15, 2025. [Online]. Available: https://platform.openai.com/docs/pricing

[28] "Claude - Models overview," Claude Docs. Accessed: Dec. 15, 2025. [Online]. Available: https://platform.claude.com/docs/en/about-claude/models/overview

[29] "xAI Models and Pricing," xAI. Accessed: Dec. 15, 2025. [Online]. Available: https://docs.x.ai/docs/models

[30] "Gemini models," Gemini API. Accessed: Dec. 15, 2025. [Online]. Available: https://ai.google.dev/gemini-api/docs/models

[31] A. Chavan, Raghav Magazine, S. Kushwaha, M. Debbah, and D. Gupta, "Faster and lighter LLMs: A survey on current challenges and way forward," *arXiv [cs.LG]*, Feb. 02, 2024. [Online]. Available: http://arxiv.org/abs/2402.01799

[32] "Streamlit," Streamlit. Accessed: Dec. 15, 2025. [Online]. Available: https://streamlit.io/
[33] "OpenAI Cookbook," Github. Accessed: Dec. 15, 2025. [Online]. Available: https://github.com/openai/openai-cookbook/blob/main/examples/api_request_parallel_processor.py
[34] "Unpaywall." Accessed: Dec. 19, 2025. [Online]. Available: https://unpaywall.org/
[35] "Directory of Open Access Journals – DOAJ." Accessed: Dec. 19, 2025. [Online]. Available: https://doaj.org/
[36] P. Windisch, F. Dennstädt, J. Weyrich, C. Schröder, D. R. Zwahlen, and R. Förster, "Reasoning models for text mining in oncology: A comparison between o1 preview, GPT-4o, and GPT-5 at different reasoning levels," *JCO Clin. Cancer Inform.*, vol. 9, no. 9, p. e2400311, Nov. 2025, doi: 10.1200/CCI-24-00311.